\documentclass[times,3p,preprint,review,12pt]{elsarticle}

\usepackage{amsmath}
\usepackage{amsfonts}
\usepackage{amssymb}
\usepackage{upgreek}
\usepackage{kotex, color}

\usepackage{lineno}
\usepackage{hyperref}
\usepackage{multicol}
\usepackage{multirow}
\biboptions{sort&compress} 

\usepackage{lipsum}

\hypersetup{pdfauthor=author}

\journal{}

\begin{document}

\begin{frontmatter}



\title{Bifacial near-field thermophotovoltaic converter with transparent intermediate substrate}



\author[a,b]{Minwoo Choi}
\author[c]{Jaeman Song\corref{*}}
\author[a,b]{Bong Jae Lee\corref{*}}
\affiliation[a]{organization={Department of Mechanical Engineering, KAIST},
            city={Daejeon 34141},
            country={South Korea}}
\affiliation[b]{organization={Center for Extreme Thermal Physics and Manufacturing, KAIST},
            city={Daejeon 34141},
            country={South Korea}}
\affiliation[c]{organization={Department of Mechanical Engineering, Kyung Hee University},
            city={Yongin 17104},
            country={South Korea}}
\cortext[*]{bongjae.lee@kaist.ac.kr, jaemansong@khu.ac.kr}

\begin{abstract}
Thermophotovoltaic (TPV) converters are capable of generating electrical energy from infrared radiation emitted by an emitter powered by waste heat or solar energy. Key performance metrics for TPV converters are power output density (POD), which represents the electrical energy per unit area of the photovoltaic (PV) cell, and converter efficiency (CE), which indicates the proportion of radiative energy converted into electrical energy. A common method to significantly enhance POD is maintaining a sub-micron vacuum gap between the emitter and PV cell to leverage the near-field thermal radiation. On the other hand, bifacial TPV conversion, operating in the far-field regime, has been proposed to enhance CE by efficiently recycling the sub-bandgap energy radiation. However, bifacial TPV converters face a challenge in cooling the PV cell because the excess heat should be transferred in the lateral direction to side-edge cooling channels. Therefore, careful thermal engineering and management are required when employing near-field thermal radiation effects on bifacial TPV converters.  In this study, we propose a bifacial near-field TPV (NF-TPV) converter that incorporates intrinsic Si intermediate layers, aiming to enhance both POD and CE. Si layers cover both sides of the PV cell to play a crucial role in PV cell cooling while addressing surface mode photonic loss in NF-TPV converters. We comprehensively analyze the influences of design parameters for a practical design of the bifacial NF-TPV converter. Our results demonstrate that a single-junction InAs cell can harvest 4.38 W/cm$^2$ of electrical energy with 27.2\% CE from 1000 K graphite emitters at a 100 nm vacuum gap. Despite the challenge in the cooing, our bifacial NF-TPV converter demonstrates 2.4 times larger POD with 2.7\% larger CE compared to conventional NF-TPV converters.
\end{abstract}



\begin{keyword}
Thermophotovoltaics \sep Near-field thermal radiation \sep Bifacial thermophotovoltaics
\end{keyword}
\end{frontmatter}



\section{Introduction}

Thermophotovoltaic (TPV) converters can directly produce electrical energy from the radiative energy emitted by a hot source without moving parts \cite{zhou2015prospects, burger2020present, datas2021TPV}. Driven by waste heat \cite{zhao2017high, lu2018inas, licht2019review}, solar energy \cite{wurfel1980upper, wang2019solar, jia2022parametrical, shan2022comparison} or thermal batteries \cite{datas2022latent}, TPV converters have potential for eco-friendly and high-performance energy converter. When radiation with energy larger than the bandgap energy of the PV cell (i.e., above-bandgap radiation) is transferred, electrical power can be generated. Two crucial performance metrics of TPV converters are the converter efficiency (CE), which indicates the proportion of transferred radiative energy that is converted into electrical power, and the power output density (POD), which represents the generated electrical power per unit area of the photovoltaic (PV) cell. Various theoretical and experimental efforts have been made to improve CE and POD of TPV converters.

One approach to significantly enhance POD is by maintaining a nano-scale gap between the emitter and PV cell to harness near-field thermal radiation \cite{park2008performance, francoeur2011thermal, liao2017parametric, shan2021parametric, yang2022performance, song2022multi, mittapally2023near}. The emitted radiative heat flux in the far field is limited to blackbody radiation by Planck's law. When the distance between the emitter and PV cell is smaller than the characteristic wavelength of the thermal radiation, previously untransmitted radiation modes (i.e., evanescent modes) can now be transmitted. The heat flux by near-field radiation increases inversely proportional to the square of the sub-wavelength-sized gap, making it possible to largely boost the POD \cite{park2013fundamentals}. In an experimental demonstration of near-field TPV (NF-TPV) converters, advancements in experimental techniques have led to the achievement of POD exceeding the blackbody limit in an mm-scale device \cite{inoue2021integrated, song2022modeling}.

Key strategies for achieving high CE in TPV conversion include minimizing photonic loss and electrical loss. Photonic loss is composed of thermalization loss caused by excess energy of above-bandgap radiation, absorption of radiation with energy lower than the bandgap energy of the PV cell (i.e., sub-bandgap radiation), and absorption of radiation by the substrate and reflector that lies behind the PV cell \cite{song2022modeling}. The latter two are jointly referred to as parasitic absorption loss. Electrical loss arises from the non-radiative recombination of electrons and holes in the PV cell, which intensifies as the temperature and thickness of the PV cell increase. The excess heat by unconverted radiation burdens the cooling component, which provides an additional reason to reduce photonic and electrical losses.

For the reduction of photonic loss, energy-selective absorption of radiation at PV cells is crucial. Selective emission from emitters via photonic engineering was a popular method \cite{liu2011taming, cho2019optical}, where the emitter often became costly and concerns for the thermal stability of photonic structure emerged. Alternatively, cheap and thermally robust emitters that exhibit broad-band emission can achieve high CE when absorption of sub-bandgap radiation at PV cell is mostly suppressed \cite{omair2019ultraefficient, fan2020near, song2022multi}. By combining thin-film PV cells with backside reflectors (BSR), the emitted sub-bandgap radiation can be transmitted through the PV cell without absorption and reflected back to the emitter. Due to multiple reflections within the PV cell boundaries, the high reflectivity of BSR is necessary to enhance the ratio of recycled sub-bandgap radiation. CE of 29.1\% has been experimentally achieved using Au reflector \cite{omair2019ultraefficient}. Advanced MEMS structure incorporating an air gap between the PV cell and the Au reflector enabled the portion of recycled sub-bandgap radiation as high as 98.5\% to achieve CE exceeding 40\% \cite{fan2020near}. The BSR-based approach has been popularly applied to NF-TPV conversion in both theoretical \cite{bright2014performance, song2022comprehensive, song2022multi} and experimental \cite{mittapally2021near, inoue2021integrated, song2022thermophotovoltaic} works, in order to explore possibilities of extremely high POD together with enhanced CE. Since the emission of sub-bandgap radiation was substantially enhanced by photon tunneling, absorption of sub-bandgap radiation was not negligible in experimental work even though the thin-film PV cell was utilized, and achieved CE was limited \cite{mittapally2021near}. Furthermore, the highly-reflective air-gap reflector would be challenging to apply to NF-TPV converters due to the structural bowing of the PV cell \cite{fan2020near}. A more efficient method of recycling sub-bandgap radiation must be applied if the performance of the NF-TPV converter is to be further improved.

Recently, an improvement for the sub-bandgap photon recycling method called bifacial TPV conversion has been suggested \cite{burger2022semitransparent,datas2023bifacial}. This method involves placing emitters on both sides of the PV cell so that emitted sub-bandgap photons can be absorbed by the emitter on the opposite side without the need for reflection at BSR. With the given performance of the PV cell, bifacial TPV conversion exhibits superior efficiency compared to the BSR-based approach. At the same time, POD in bifacial TPV converters could potentially be twice as high or even greater, owing to the implementation of a dual emitter and the reduction in photonic losses. Hence, integrating bifacial TPV conversion with NF-TPV conversion has the potential to achieve unprecedented POD, along with improved CE compared to the conventional method of NF-TPV conversion utilizing backside reflectors. A comprehensive exploration of the potential of a bifacial near-field TPV converter is yet to be provided.

PV cells in TPV converters need to be kept at low temperatures to suppress electrical loss. In BSR-Based TPV converters, PV cells can be directly cooled down from the bottom side of the BSR. On the other hand, bifacial TPV converters require excess heat to be transferred in a lateral direction toward the cooling channels at two side edges via conduction \cite{datas2023bifacial}. As thin-film PV cells have an inadequate cross-sectional area for sufficient conduction transparent, intermediate layers attached to the PV cell become necessary in bifacial TPV converters. Even with the effort, the PV cell temperature rise and corresponding performance loss could be substantial in the far-field regime. In comparison, the bifacial NF-TPV converter is expected to exhibit significantly larger excess heat that needs to be dissipated. We explore for the first time the potential of bifacial NF-TPV converters incorporating transparent intermediate layers by careful examination of performance through realistic calculation.

In this study, we propose a practical design of bifacial NF-TPV conversion utilizing a bulk graphite emitter, indium arsenide (InAs) PV cell, and intrinsic Si intermediate layers and perform a comprehensive analysis of its performance. We apply intrinsic Si intermediate layers on both sides of the PV cell for the PV cell cooling and suppression of the absorption of low-energy surface mode radiation \cite{inoue2021integrated}. Here, the thickness of the Si intermediate layer is a crucial design parameter that differently affects various photonic losses and the PV cell's cooling performance. To iteratively solve PV cell temperature and current density profile in the lateral direction, an opto-electro-thermal calculation framework is developed which integrates fluctuation electrodynamics, detailed balance analysis, and 1-D heat diffusion problem. The effect of vacuum gap size, emitter temperature, and lateral size of the converter on its POD and CE are comprehensively investigated with consideration of realistic design factors, such as edge cooling heat flux and view factor between two emitters. We demonstrate that a practical design for a bifacial NF-TPV converter can be suggested by adjusting the thickness of the Si intermediate layer and the converter's lateral size. Our research reveals the potential of the bifacial NF-TPV converter for cost-effective high-performance TPV conversion, suggesting the same POD from emitters with much lower temperatures compared to conventional BSR-based NF-TPV converters.

\section{Methods}

\subsection{Design fundamentals and roles of Si intermediate layer}

\begin{figure}[!t]
    \centering
    \includegraphics[width=0.45\textwidth]{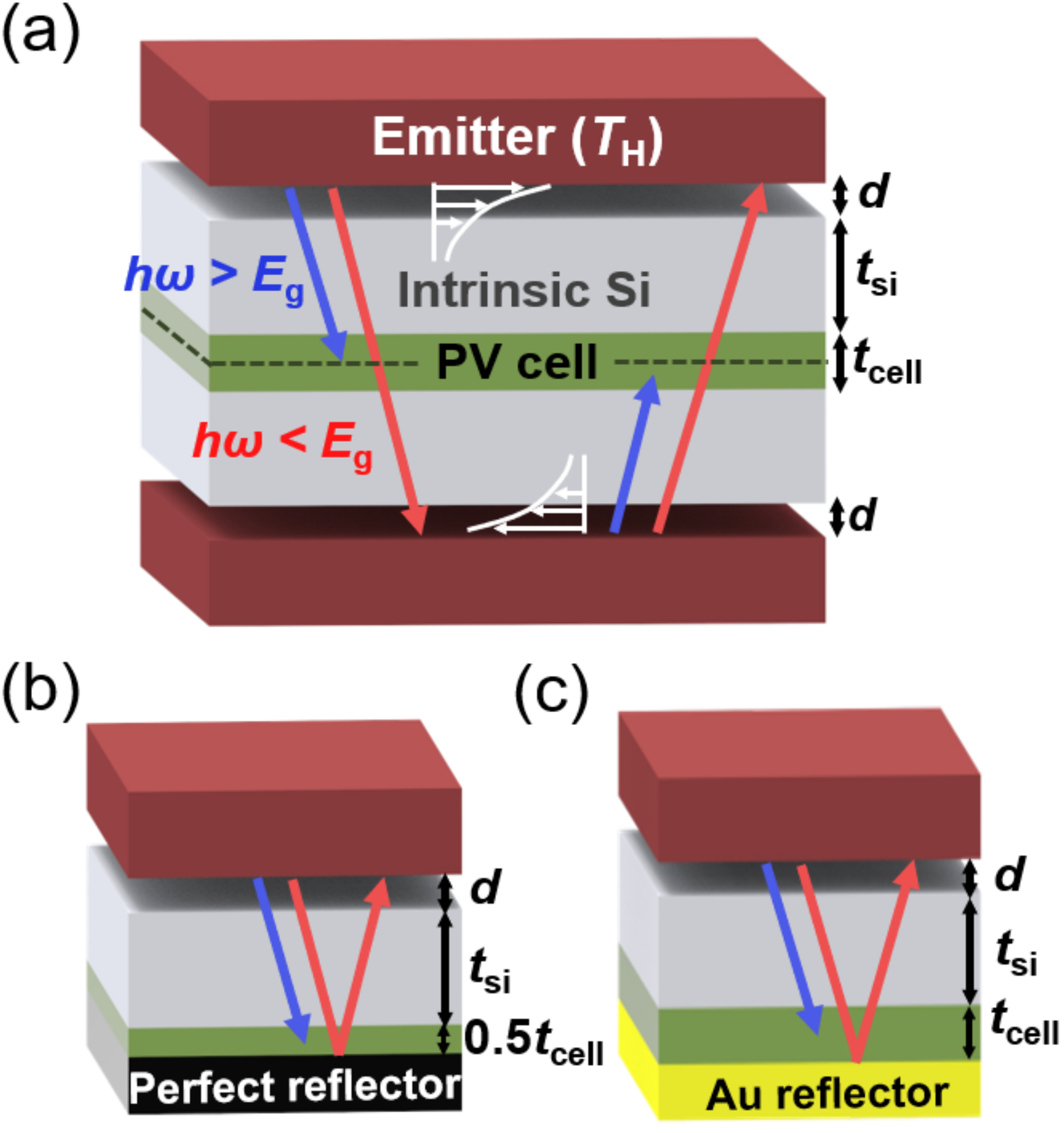}
    \caption{(a) Schematic of the suggested bifacial NF-TPV converter, (b) Schematic of the optically equivalent TPV converter with perfect reflector introduced for the efficient calculation of absorbed radiative heat flux of bifacial NF-TPV converter, (c) Schematic of the NF-TPV converter with Au reflector for performance comparison.}
    \label{fig:1}
\end{figure}

Fig.\ \ref{fig:1}(a) illustrates the schematic of the proposed bifacial near-field thermophotovoltaic (NF-TPV) converter. Two graphite emitters at a constant temperature $T_{\text{H}}$ emit radiative heat flux while maintaining a vacuum gap distance $d$ with the absorbing composite. The absorbing composite has a sandwich-like structure, with an InAs PV cell placed between two intrinsic Si intermediate layers each with thickness $t_{\text{si}}$. Both the emitters and the absorbing composite have lateral size $L$. The intrinsic Si maintains high transparency in energy regimes where most radiation is transferred, with minimal radiation absorption by itself unless $t_{\text{si}}$ becomes too thick. When sub-bandgap radiation penetrates the absorbing composite and is absorbed by the emitter on the opposite side, the emitted radiation is recycled without loss. As the converter structure is vertically symmetric, the spectral absorbed heat flux to the absorbing composite can be analyzed with an optically equivalent structure with a perfect reflector illustrated in Fig.\ \ref{fig:1}(b). After obtaining spectral absorbed heat flux to Si intermediate layer and InAs cell in the equivalent structure, multiplying them by 2 gives the actual absorption for the bifacial NF-TPV converter. Without an effort to enhance the reflectivity of BSR, extremely efficient sub-bandgap photon recycling is possible which has been a desired feature for NF-TPV converters \cite{desutter2017external,datas2023bifacial}.

We utilize bulk graphite emitters that can be easily processed with high surface quality and also exhibit strong broadband emissions in the infrared energy regime. The emitter is regarded as semi-infinite, i.e., significantly thicker than the penetration depth of radiation. The PV cell thickness $t_{\text{cell}}$ is set to 200 nm to make most sub-bandgap radiation penetrate while sufficiently absorbing the above-bandgap radiation. We use InAs PV cell with a bandgap energy of 0.355 eV at the temperature of 300 K. The bandgap energy is sufficiently low to absorb a significant amount of radiation, but high enough to suppress the thermalization loss at a range of $T_{\text{H}}$ we focused on (800 $\sim$ 1500 K). The absorbed above-bandgap radiation in the PV cell generates an electron-hole pair (EHP) from a photon. Among the photogenerated EHPs, the part not diminished as non-radiative recombination (i.e., electrical loss) is collected as the photocurrent. Non-radiative recombination escalates as the bias voltage applied to the PV cell increases. POD is obtained where the multiplication of photocurrent per unit cell area (i.e., photocurrent density) and the bias voltage is maximized, while CE is obtained by dividing the POD by the total absorption by the absorbing composite. In bifacial NF-TPV converters, dual emitters lead to approximately twice the absorbed radiative heat flux compared to a TPV converter with Au reflector illustrated in Fig.\ \ref{fig:1}(c). Consequently, the photogeneration rate also increases at a similar rate. However, the magnitude of non-radiative recombination loss remains unchanged when $t_{\text{cell}}$, $T_{\text{L}}$ and electrical properties of PV cell are the same for the two NF-TPV converters. Thus, the bifacial TPV converter can induce an increase in open-circuit voltage compared to BSR-based converters, leading to a larger than twofold enhancement in POD.

In the bifacial NF-TPV converter design, intrinsic Si intermediate layers play multiple roles simultaneously. The first role is offering physical support and preventing deformation by external forces. The second role is assisting in the cooling of PV cells. Excess heat from photonic and electrical losses needs to escape through the cooling channels on each side edge of the absorbing composite. The cooling channels maintain a constant temperature $T_{\text{cool}}$, while a lateral temperature profile $T_{\text{L}}$($x$), arises in the absorbing composite. The temperature rise of PV cells causes a large enhancement in non-radiative recombination and has to be vigorously suppressed. The Si intermediate layer increases the conduction cross-sectional area while having a thermal conductivity level similar to that of metals (150 W/m$\cdot$K). Therefore, the temperature rise can be effectively suppressed as the thicker Si intermediate layer is applied. Since the excess heat in the unit area of the PV cell proportionally rises with POD at a similar CE level, the cooling load of bifacial NF-TPV converters is much higher than in bifacial far-field TPV converters (\cite{datas2023bifacial}). However, increasing $t_{\text{si}}$ as large as to suffice such an intense cooling load may result in non-negligible radiation absorption by Si intermediate layer or geometrically impractical structure. Therefore, careful control of $t_{\text{si}}$ is a crucial task that will be addressed throughout this paper. Even with the same $t_{\text{si}}$, longer lateral size results in higher maximum and lateral average in $T_{\text{L}}$($x$). Joint consideration of $L$ and $t_{\text{si}}$ for CE and POD will be discussed in the final part of this paper. In lieu of conductive heat dissipation toward cooling channels at the lateral edges, convective cooling in the vicinity of the PV cell surface could be employed when internal micro-channels are incorporated within the Silicon intermediate layers. Nevertheless, this approach will not be taken into consideration in the current study.

The third role of the Si intermediate layer is the reduction of surface mode photonic loss in NF-TPV conversion. Near-field thermal radiation includes frustrated mode that propagates in the Si intermediate layer and can be transferred to the PV cell, and surface mode that can only exist near the vacuum gap. If the InAs cell faces the graphite emitter without the Si intermediate layer, surface phonon polaritons (SPhPs) can be excited which can transfer low-energy radiation to the InAs cell and cause photonic loss \cite{inoue2018near, inoue2021integrated}. On the other hand, intrinsic Si does not support SPhP and can be safely placed near the vacuum gap. Unlike the surface mode, photon tunneling in frustrated modes occurs across entire spectral regimes, amplifying the above-bandgap radiation to increase the photogeneration rate and POD. Since the refractive index of intrinsic Si in the above-bandgap regime is similar to that of InAs, the frustrated mode radiation can be safely transported to the InAs cell \cite{inoue2018near}.

The fourth role is assisting in the maintenance of a nano-scale gap by allowing for the fabrication of spacers. Maintaining the nano-scale gap between the high-temperature emitter and PV cell over the mm-scale PV cell area is an extremely challenging task. Experimental demonstration of NF-TPV converter has been achieved over smaller emitter areas \cite{fiorino2018nanogap, mittapally2021near, song2022thermophotovoltaic} or with modified emitter-PV cell geometries other than the plane-plane one \cite{lucchesi2021near}. In the only implementation of mm-scale plane-plane NF-TPV converter, spacers were placed at regular intervals in the vacuum gap to prevent contact between the emitter and absorbing composite \cite{inoue2019one}. In the work, a Si intermediate layer was placed on the PV cell to face emitters. An array of spacers were fabricated on top of the Si intermediate layer to guarantee the maintenance of the nano-scale gap with the emitter. To allow for spacers much longer than the vacuum gap size for sufficient thermal resistance, each spacer was buried inside a pit fabricated on the Si intermediate layer. By the Si intermediate layer, the PV cell could also be protected from direct contact with the high-temperature emitter. Based on previous works, implementation of a practical NF-TPV converter are likely possible when the Si intermediate layer faces the emitter. Consequently, Si intermediate layers enable the experimental implementation of a bifacial NF-TPV converter with two nano-scale gaps.

\subsection{Opto-electro-thermal calculation framework}

The performance calculation of bifacial NF-TPV involves deriving the photocurrent density-bias voltage relation ($J$-$V$ relation) for the PV cell and determining the maximum power point where the product of $J$ and $V$ is maximized to calculate the power output density (POD) and converter efficiency (CE). We adopt an optical model to describe the net spectral radiative heat flux $q(\omega)$ absorbed by the PV cell and Si intermediate layers, an electrical model to derive the photocurrent density $J$ collected at the PV cell, and a thermal model to relate the lateral temperature profile $T_{\text{L}}$($x$) of the absorbing composite to the excess heat from it. Since net spectral heat flux and electrical loss in PV cell varies with $T_{\text{L}}$($x$), the numerical calculation is required for the derivation of photocurrent density profile $J$($x$). Representative $J$ for a certain $V$ is given by a spatial average in $x$. Since the optical model, electrical model, and thermal model are coupled via $T_{\text{L}}$($x$), we develop an opto-electro-thermal calculation algorithm that iteratively finds photocurrent density profile $J$($x$) at a certain $V$.

First, we describe the optical model to find the absorbed heat flux. The bifacial NF-TPV converter has a symmetric structure based on the center of the PV cell. Therefore, the radiative heat flux can be calculated by replacing it with an optically equivalent structure shown in Fig.\ \ref{fig:1}(b). In the equivalent structure, the thickness of the PV cell is reduced to half of the original thickness, and a perfect reflector is placed at the backside of the PV cell. The spectral absorbed radiative heat flux for each layer in the equivalent structure is multiplied by two to obtain the results in the actual bifacial case. In the equivalent structure, the absorbing composite is composed of a perfect reflector, InAs cell, and intrinsic Si where a constant temperature $T_{\text{L}}$ is assumed. A vacuum gap $d$ exists between absorbing composite and graphite emitter at temperature $T_{\text{H}}$. Assuming a 1-D layered geometry, the total net absorbed radiative heat flux $Q_{j}$ by layer $j$ in absorbing composite is calculated from \cite{francoeur2009solution}
\begin{equation}
Q_{j} = \int_{0}^{\infty} \hbar \omega [\Theta(\omega,T_{\text{H}},0)-\Theta(\omega,T_{\text{L}},\mu)] \Phi_{mj}(\omega) A_{\text{Bi}} d\omega
\label{eq:1}
\end{equation} where $\omega$ is angular frequency. The calculation range for $\omega$ is from 3.4 $\times$10$^{13}$ rad/s to 4 $\times$10$^{15}$ rad/s. $\Theta$($\omega$,$T$,$\mu$) represents the mean energy of the Planck oscillator, which is described as (exp\{($\hbar$$\omega$-$\mu$)/$k_{\text{B}}$$T$-1\})$^{-1}$ when photon energy $\hbar$$\omega$ is larger than the bandgap energy and (exp\{$\hbar$$\omega$/$k_{\text{B}}$$T$-1\})$^{-1}$ when $\hbar$$\omega$ is smaller than the bandgap energy. $k_{\text{B}}$ is the Boltzmann constant, $\hbar$ is the reduced Planck constant. $\mu$ is the chemical potential of the photon in the PV cell expressed as $\mu$ = $e$$V$, where $e$ is the charge of an electron, and $V$ is the bias voltage of the PV cell. $T_{\text{H}}$ and $T_{\text{L}}$ are emitter and PV cell temperatures, respectively. $\Phi_{mj}$($\omega$) represents the proportion of photons emitted from the emitter layer $m$ and absorbed in layer $j$ in the absorbing composite. It is calculated from the Weyl components of the dyadic Greens' tensors \cite{francoeur2009solution} with takes the permittivity of materials composing layers $m$ and $j$ as inputs. The perfect reflector is treated as a virtual material with infinite real and imaginary parts of permittivity. The permittivity of intrinsic Si, graphite, and Au was found in \cite{palik1998handbook}. The temperature-dependent permittivity of InAs was obtained using Adachi's model \cite{adachi1989JAP}. $A_{\text{Bi}}$ is set as 2 for the bifacial TPV converter (Fig.\ \ref{fig:1}(b)) and 1 for the TPV converter with Au reflector (Fig.\ \ref{fig:1}(c)).

Next, we describe the electrical model to find the retrieved photocurrent density $J$ at bias voltage $V$. It is assumed that the internal quantum efficiency of the PV cell is 1, meaning that every absorbed above-bandgap photon generates an EHP. Since infinite carrier mobility is assumed for thin PV cells, detailed balance analysis can be safely used. Photocurrent density $J$ can be calculated from
\begin{equation} \label{eq:2}
\begin{split}
    J(V) & = e\{G_{\text{p}} - G_{\text{r}}(V) - R_{\text{nr}}(V)\} \\
& = e\Biggr[\int_{\omega_{\text{g}}}^{\infty} \Theta(\omega,T_{\text{H}},0)\Phi_{mp}(\omega)d\omega - \int_{\omega_{\text{g}}}^{\infty} \Theta(\omega,T_{\text{L}},\mu)\Phi_{pm}(\omega)d\omega - (R_{\text{Auger}}+R_{\text{SRH}}+R_{\text{Surf}})\Biggr]
\end{split} 
\end{equation} where $G_{\text{p}}$ represents the photogeneration rate, and $G_{\text{r}}$ indicates the radiative recombination rate, which is the photo-luminescence emitted from the PV cell at bias voltage $V$. $G_{\text{p}}\--G_{\text{r}}$ refers to net photogeneration rate. $\Phi_{mp}$($\omega$) denotes the proportion of photons emitted from the emitter layer to the PV cell layer, while $\Phi_{pm}$($\omega$) denotes the proportion of photons transferred in the reverse direction via the radiative recombination. $R_{\text{nr}}$($V$) is the non-radiative recombination rate, obtained from the sum of Auger recombination, Shockley-Read-Hall (SRH) recombination, and surface recombination. Expressions for $R_{\text{Auger}}$, $R_{\text{SRH}}$, and $R_{\text{Surf}}$ together with bias-dependent carrier concentrations are driven according to \cite{song2022modeling}. Auger recombination coefficient, electron and hole trap concentrations, and SRH lifetime for InAs were found in \cite{ioffeInAs}, which were assumed constant with temperature for this work. The thickness of the PV cell used in the calculation of Auger and SRH recombination rates was 200 nm. The surface recombination velocity is assumed to be 100 m/s at both surfaces of the PV cell. The series resistance loss, which may degrade the performance of NF-TPV converters \cite{vaillon2019micron, song2022comprehensive}, was not assumed in this work. In NF-TPV converters without Si intermediate layers, the thickness of the front contact electrode is restricted by the vacuum gap size, causing a huge loss by series resistance. When using Si intermediate layer, such constraint vanishes since the Si layer can be etched by microns to provide space for sufficiently thick electrodes.

\begin{figure}[!t]
    \centering
    \includegraphics[width=0.55\textwidth]{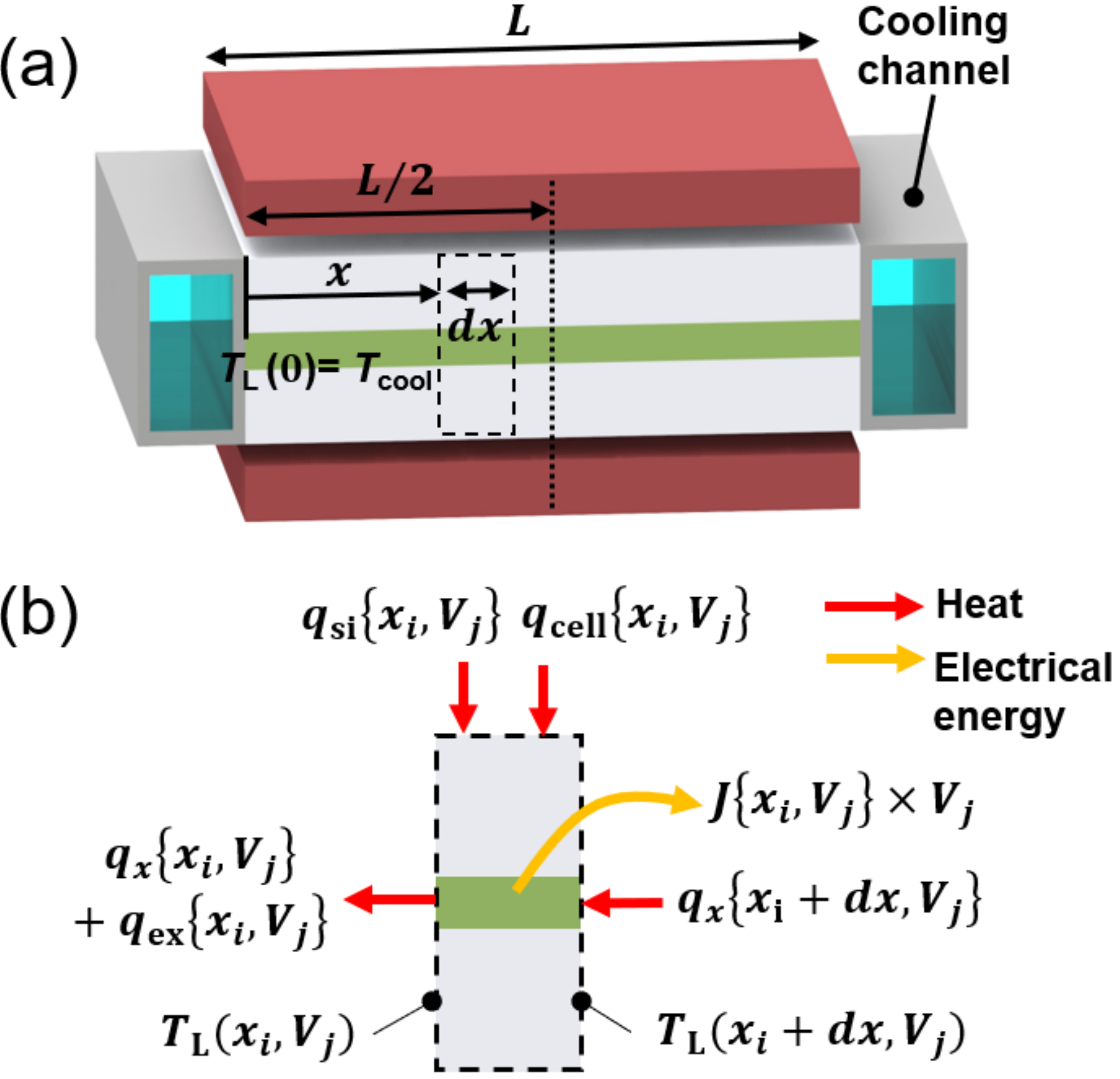}
    \caption{(a) Schematic of the bifacial NF-TPV converter with coordinates, dimensions, and control volume for the thermal modeling, (b) Energy flow diagram for the $i$-th control volume depicted in (a), at the bias voltage $V_{j}$.}
    \label{fig:2}
\end{figure}

We describe the iterative calculation algorithm for obtaining the $J$-$V$ relation with the reflection of the lateral temperature profile of the absorbing composite. Refer to the coordinates and control volume defined in Fig.\ \ref{fig:2}(a) for thermal modeling. Considering the symmetry in the lateral direction, the left half of the absorbing composite is divided into 200 control volumes of width $dx$. It is assumed that the temperature within the $i$-th control volume is at $T_{\text{L}}$($x_{i}$). The algorithm starts with the analysis for the control volume at the side edge ($x_{0}$=0) where $T_{\text{L}}(0)$ = $T_{\text{cool}}$. The bias voltage is set as a certain fixed value $V_{j}$. Absorbed radiative heat fluxes to Si intermediate layer ($q_{\text{si}}$($x_{i}$, $V_{j}$)) and PV cell ($q_{\text{cell}}$($x_{i}$, $V_{j}$)) are obtained from Eq.\ \ref{eq:1} and photocurrent density $J(x_{i}, V_{j})$ is obtained from Eq.\ \ref{eq:2}. Fig.\ \ref{fig:2}(b) describes the energy balance of the $i$-th control volume. The amount of excess heat $q_{\text{ex}}$ due to photonic and electrical losses in the control volume is calculated as
\begin{equation}
q_{\text{ex}}(x_{i}, V_{j}) = q_{\text{si}}(x_{i}, V_{j})+q_{\text{cell}}(x_{i}, V_{j})-\{J(x_{i}, V_{j})\times V_{j}\}
\label{eq:3}
\end{equation} Temperature of the control volume next to the current control volume is obtained by treating the situation as a one-dimensional heat diffusion problem, as in
\begin{equation}
T_{\text{L}}(x_{i}+dx, V_{j}) = T_{\text{L}}(x_{i},V_{j}) + \frac{q_{\text{ex}}\times dx}{k_{\text{si}}\times2t_{\text{si}}}
\label{eq:4}
\end{equation} where $k_{\text{si}}$=150 W/m$\cdot$K is thermal conductivity of intrinsic Si. The same process is repeated for the control volume next to the current control volume with $i$=$i$+1 until the center temperature $T_{\text{L}}$($x$=$L$/2, $V_{j}$) is found. Starting from the control volume at the side edge ($i$=0) again, the process is repeated with the updated temperature profile. The calculation is finished when $T_{\text{L}}$($x$=$L$/2, $V_{j}$) converges, where the relative difference between calculations in previous and current iterations are less than 10$^{-5}$. The calculation results in lateral profiles of temperature $T_{\text{L}}$($x$, $V_{j}$) and current density $J$($x$, $V_{j}$) corresponding to a certain bias voltage $V_{j}$. Representative $T_{\text{L}}$($V_{j}$) and $J$($V_{j}$) values are obtained from spatial average over $x$ from $x$ = 0 to $x$ = $L$/2. The whole procedure outlined above is repeated each for a series of $V_{j}$ values equally spaced between $V_{\text{0}}$=0 and $V_{\text{max}}$ to obtain the $J$-$V$ relation. $V_{\text{max}}$ is set as 0.381 V which is higher than the bandgap energy of InAs at 300 K. Converter efficiency is calculated from
\begin{equation}
\text{Converter efficiency} = \frac{J_{\text{mpp}}\times V_{\text{mpp}}}{q_{\text{si}}(V_{\text{mpp}})+q_{\text{cell}}(V_{\text{mpp}})} 
\label{eq:5}
\end{equation} where subscript `mpp' stands for maximum power point.

\section{Results and discussion}

\subsection{Performance enhancement mechanisms}

\begin{figure}[!t]
    \centering
    \includegraphics[width=0.55\textwidth]{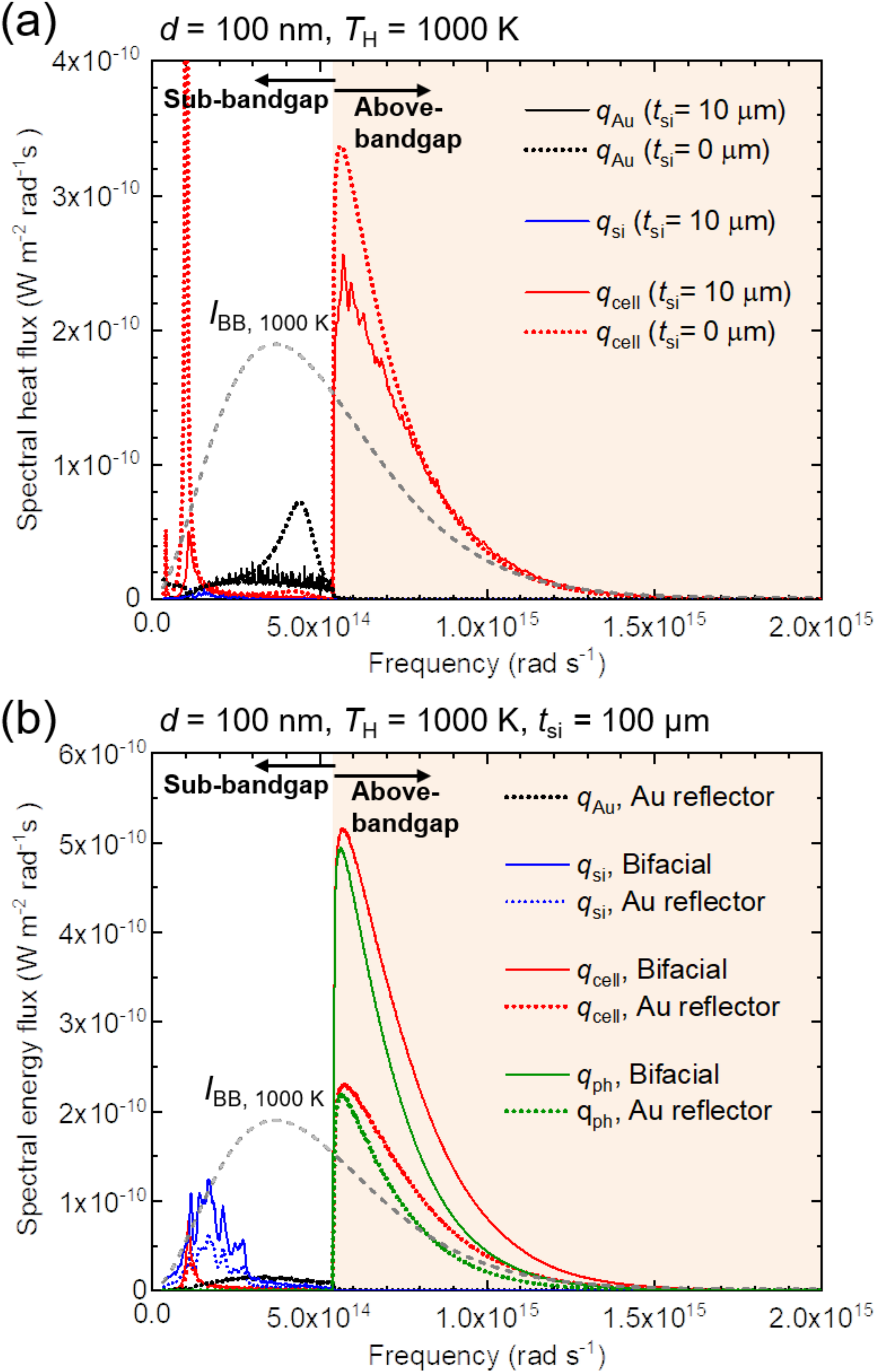}
    \caption{(a) Spectral absorbed heat fluxes to the Si intermediate layer $q_{\text{si}}$($\omega$), PV cell $q_{\text{cell}}$($\omega$), and Au reflector $q_{\text{Au}}$($\omega$) in NF-TPV converter with Au reflector and 100 nm vacuum gap, 1000 K graphite emitter, and 200-nm-thick InAs cell at 300 K. Both the cases with and without 10-$\mu$m-thick Si intermediate layer are displayed together, (b) Spectral absorbed heat flux and spectral net photogeneration heat flux of two types of NF-TPV converters (bifacial and with Au reflector) at $V$ = 0 and at constant PV cell temperature of 300 K.}
    \label{fig:3}
\end{figure}
Before we investigate the bifacial TPV converter shown in Fig.\ \ref{fig:1}(a), we show that Si intermediate layer between the vacuum gap and the PV cell is a preferred feature in NF-TPV converters due to its role in photonic loss reduction. To this end, we analyze the NF-TPV converter with Au reflector (Fig.\ \ref{fig:1}(c)), which is generally designed without a Si intermediate layer since mechanical support is not essential, in contrast to bifacial TPV converters. For the NF-TPV converter with an Au backside reflector, the layer-by-layer spectral net radiative heat fluxes calculated from Eq.\ \ref{eq:1} are shown in Fig.\ \ref{fig:3}(a). Calculation results for converters with ($t_{\text{si}}$ = 10 $\mu$m) and without ($t_{\text{si}}$ = 0 $\mu$m) Si intermediate layer are displayed together and the above-bandgap and sub-bandgap frequency regimes are indicated. In both cases with different $t_{\text{si}}$ values, $q_{\text{cell}}$($\omega$) exceeds the blackbody radiation at 1000 K ($I_{\text{BB}}$) in the above-bandgap regime.

When the Si intermediate layer is absent, a sharp peak in $q_{\text{cell}}$($\omega$) appears in the sub-bandgap regime. According to the Lorentz model of permittivity, transverse optical phonons in the InAs cell result in a negative real part of permittivity in the far-infrared regime for wavelengths longer than 17.1 $\mu$m. This negative permittivity leads to the excitation of SPhP, enhancing the absorbed radiation in InAs. The wavelengths with negative real permittivity coincide with the frequency regime of $\omega$ $\leq$ 1.1$\times$10$^{14}$ rad/s, where the sub-bandgap absorption peak emerges, suggesting that the peak is likely due to surface mode radiation induced by SPhP.

When a 10-$\mu$m-thick Si intermediate layer is added between the vacuum gap and InAs cell, surface modes can no longer be excited in InAs, resulting in attenuation of the sharp peak in $q_{\text{cell}}$($\omega$). Frustrated modes can still be transferred to the InAs cell through the Si intermediate layer, still exhibiting significantly enhanced above-bandgap absorption compared to the far-field case. Although the decrease in above-bandgap radiation, the effect of the surface mode suppression Additionally, absorption into the Si intermediate layer remains minimal due to its low absorption coefficient and the absence of surface modes, as intrinsic Si exhibits a positive real part of permittivity across all frequencies. Consequently, the converter efficiency is improved from 20.9\% to 23.6\% by incorporating the Si intermediate layer. With a smaller vacuum gap of $d$ = 10 nm, the introduction of the Si intermediate layer can enhance the CE by over 10\%. In bifacial TPV converters where Si intermediate layers become essential components, CE can be enhanced similarly by photonic loss reduction.

In Fig.\ \ref{fig:3}(b), spectral net radiative heat fluxes $q_{\text{Au}}$($\omega$), $q_{\text{si}}$($\omega$), $q_{\text{cell}}$($\omega$), and spectral net photogeneration energy flux $q_{\text{ph}}$($\omega$) are denoted for bifacial NF-TPV converter and NF-TPV converter with Au reflector. The bias voltage of 0 V was used for the calculation result in Fig.\ \ref{fig:3}. $q_{\text{ph}}$($\omega$)=$q_{\text{cell}}$($\omega$)$\times$\{$\omega_{\text{g}}$/$\omega$\} indicates the portion of absorbed heat flux contributed to the generation of EHP. Accordingly, total net photogeneration energy flux $Q_{\text{ph}}$ = $\int_{\omega_{\text{g}}}^{\infty}$$q_{\text{ph}}$($\omega$)d$\omega$ = $\hbar\omega$$_{\text{g}}$($G_{\text{p}}\--G_{\text{r}}$), which is represented by the area under the green curve in Fig.\ \ref{fig:3}(b), is equal to the photocurrent of the TPV converter when there is no non-radiative recombination. Different from Fig.\ \ref{fig:3}(a), the thickness of the Si intermediate layer was increased to 100 $\mu$m for sufficient conduction cross-sectional area in bifacial TPV converters. Although $q_{\text{si}}$($\omega$) becomes not negligible due to the thickness increase, it turns out to be beneficial when the PV cell temperature increase is evaluated, as will be discussed later.

Comparing the bifacial TPV with $d$ = 100 nm (near-field) to the case with $d$ = 10 μm (far-field, where only propagating mode radiation contributes to radiative heat transfer), $Q_{\text{ph}}$ is enhanced by 9.34 times due to the additional above-bandgap radiation absorbed in the InAs cell, which is transferred through the Si intermediate layer. When comparing the bifacial NF-TPV converter to the NF-TPV converter with Au reflector, approximately a two-fold enhancement of spectral absorbed heat fluxes in the InAs cell and Si intermediate layer is achieved, primarily due to the introduction of a second emitter. Consequently, the $Q_{\text{ph}}$ in the bifacial NF-TPV converter is 2.18 times larger. 

The proportion of $Q_{\text{ph}}$ to the total absorbed radiation in absorbing composite is compared for the bifacial NF-TPV converter and the NF-TPV converter with Au reflector. It directly represents CE with electrical loss disregarded where only photonic loss terms (i.e., thermalization loss and parasitic absorption loss) are put into account. In NF-TPV conversion, by altering the sub-bandgap radiation recycling method from using an Au reflector to employing bifacial TPV conversion, the proportion of $Q_{\text{ph}}$ is increased from 61.1\% to 66.4\%. This enhancement primarily results from the elimination of radiation absorbed by the Au reflector, which accounts for 5.9\% of the total absorbed radiation in the absorbing composite. Regarding the proportion of $Q_{\text{ph}}$ turns into CE when non-radiative recombination is additionally considered, we can discuss that the CE of the NF-TPV converter is enhanced by adopting bifacial TPV conversion.

\begin{figure}[!t]
    \centering
    \includegraphics[width=0.65\textwidth]{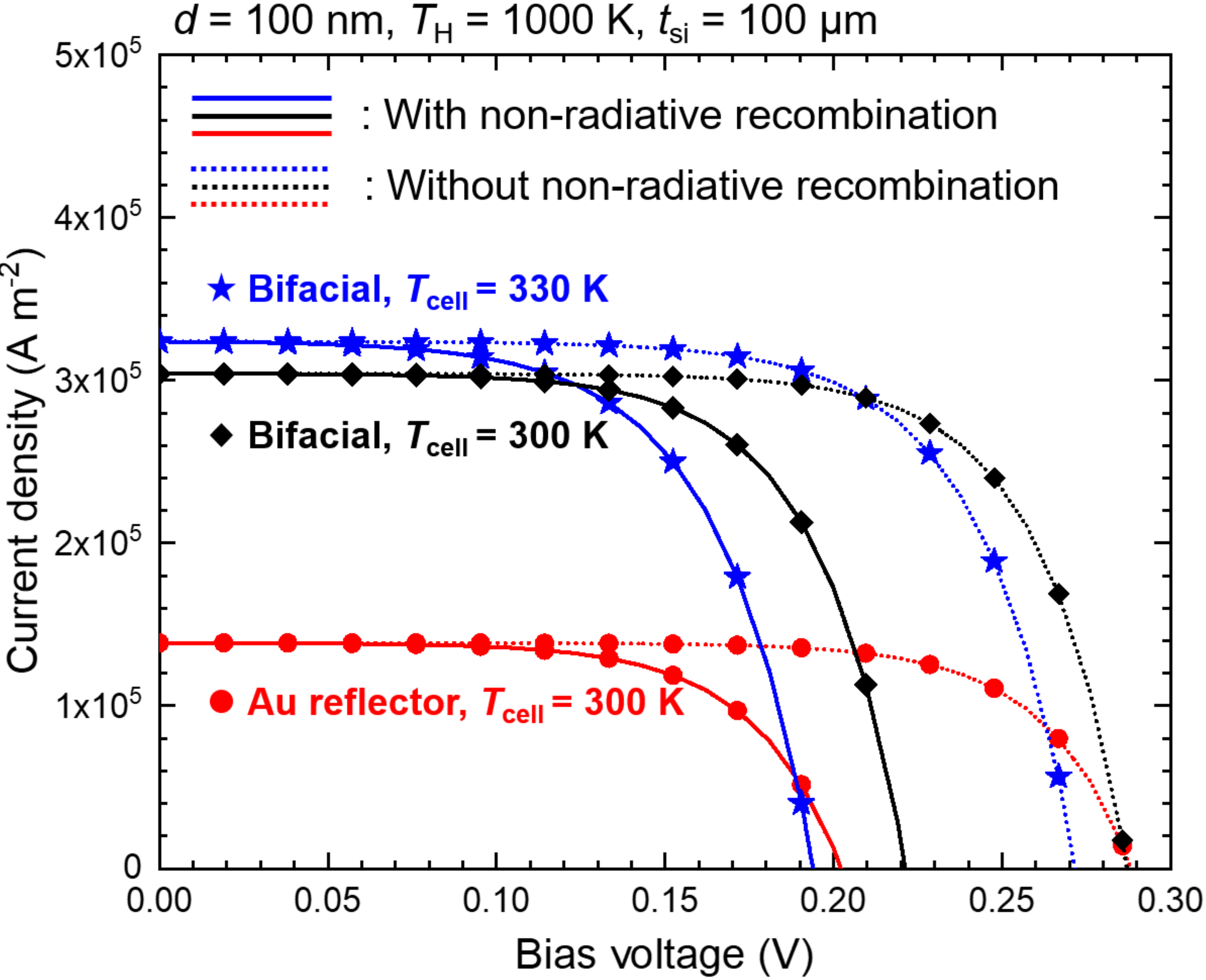}
    \caption{$J$-$V$ relations of bifacial NF-TPV converter and NF-TPV converter with Au reflector, calculated at constant PV cell temperatures based on Eqs.\ \ref{eq:1}, \ref{eq:2}. $J$-$V$ relation without non-radiative recombination denotes the photogeneration rate.}
    \label{fig:4}
\end{figure}

To understand the impact of non-radiative recombination, the current density-voltage relations of the bifacial NF-TPV converter and the NF-TPV converter with Au reflector are examined (see Fig. \ref{fig:4}). For both NF-TPV converters, Si intermediate layers with thickness 100 $\mu$m were applied. With consideration of non-radiative recombination that increases with bias voltage, the power output density (i.e., max($J$ $\times$ $V$)) at the maximum power point is 2.56 times larger for bifacial NF-TPV converter compared to that of NF-TPV converter with Au reflector. In comparison, the photogeneration rate at $V$ = 0 was only 2.18 times larger. The additional performance enhancement comes from the enhanced bias voltage at the maximum power point from 0.152 V to 0.171 V.

Bias voltage achievable at a certain photocurrent density is enhanced in bifacial NF-TPV converters. It is because only the net photogeneration rate is enhanced while the non-radiative recombination rate $R$($V$) is kept constant. In the $J$-$V$ relation without $J_{\text{nr}}$($V$) = $q$$\times$$R$($V$), voltage achieved at $J$ = 0 (i.e., open circuit voltage) is the same for two NF-TPV converters. Additionally considered $J_{\text{nr}}$($V$) is the same for two NF-TPV converters because it is determined only from the common electrical values and $V$. As seen in Fig.\ \ref{fig:4}, incorporation of the same $J_{\text{nr}}$($V$) causes a smaller voltage drop when the original net photogeneration density is higher. Accordingly, the open circuit voltage of the bifacial NF-TPV converter becomes larger than that of the NF-TPV converter with Au reflector, which explains its larger bias voltage at the maximum power point. In the case of a bifacial far-field TPV converter, the open circuit voltage is significantly reduced compared to a bifacial NF-TPV converter, primarily due to a 9.34 times smaller net photogeneration rate. Consequently, this results in a 14.0 times lower POD.

As seen in Fig.\ \ref{fig:3} and Fig.\ \ref{fig:4}, the bifacial NF-TPV converter consistently shows superior performance compared to the NF-TPV converter with Au reflector when the PV cell temperature $T_{\text{cell}}$ is kept constant. However, there inevitably exists a temperature rise in bifacial TPV converters because the absorbing composite is cooled from its two side edges. The effect of PV cell temperature rise on its $J$-$V$ relation is plotted in Fig.\ \ref{fig:4}. When $T_{\text{cell}}$ rises to 330 K, the short circuit current is increased while the open circuit voltage is reduced. Short circuit current is enhanced because the net photogeneration rate is increased due to larger above-bandgap heat flux absorbed in the InAs cell. The reason for the larger absorbed heat flux is that the bandgap energy of InAs is lowered from 0.356 eV to 0.348 eV referring to Varshini's empirical relation \cite{varshni1967temperature, ioffeInAs}. Nevertheless, the non-radiative recombination rate accelerates sharply with $T_{\text{cell}}$, effectively overshadowing the increase in the net photogeneration rate at elevated bias voltages. As a result, open circuit voltage reduction by the non-radiative recombination gets larger, resulting in much reduced $V_{\text{oc}}$ even smaller than that of NF-TPV converter with Au reflector when $T_{\text{cell}}$ is 300 K. Accordingly, by the cell temperature rise by 30 K, POD drops from 4.47 W/cm$^2$ to 3.87 W/cm$^2$ in bifacial NF-TPV converter. Because of largely increased electrical loss, CE decreases from 27.7\% to 22.7\% by 5\% point. CE of the NF-TPV converter with Au reflector is also 22.7\%. Therefore, the benefit in CE is canceled out when the PV cell temperature rises in bifacial NF-TPV beyond a certain level. In the design process bifacial NF-TPV converter, design parameters that affect the temperature rise, such as silicon thickness $t_{\text{si}}$, vacuum gap $d$, emitter temperature $T_{\text{H}}$, and lateral size $L$ need to be carefully tailored to harness large POD with increased CE.

\subsection{Effect of vacuum gap distance and Si intermediate layer thickness}

\begin{figure}[!t]
    \centering
    \includegraphics[width=0.5\textwidth]{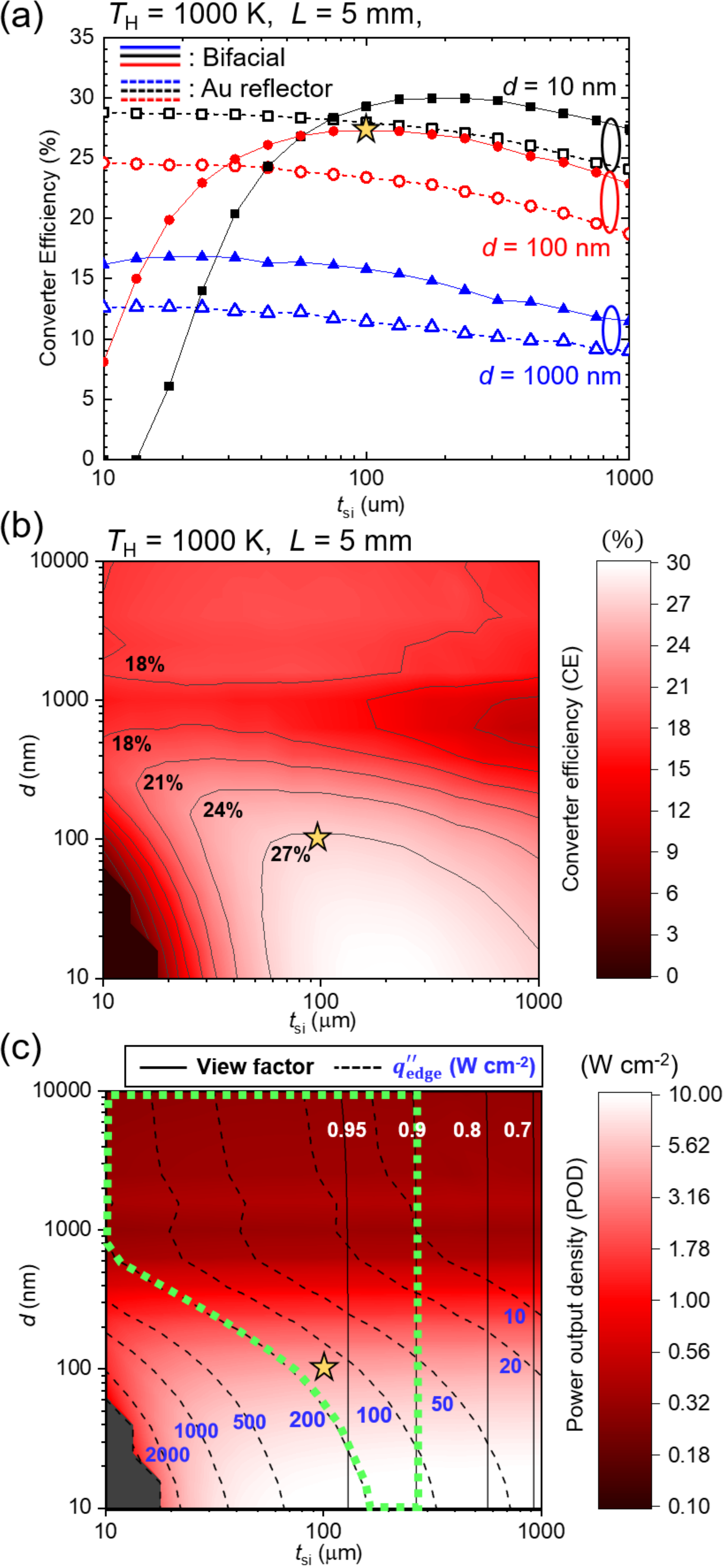}
    \caption{(a) Converter efficiency (CE) of the bifacial NF-TPV converter and the NF-TPV converter with Au reflector with respect to Si intermediate layer thickness $t_{\text{si}}$ at three different vacuum gap distances $d$, (b) CE of the bifacial NF-TPV converter with respect to $t_{\text{si}}$ and $d$, (c) Power output density of the TPV converter with respect to $t_{\text{si}}$ and $d$ together with view factor between two emitters and edge cooling heat flux.}
    \label{fig:5}
\end{figure}

In Fig.\ \ref{fig:5}, the effect of vacuum gap $d$ and $t_{\text{si}}$ to CE and POD of bifacial NF-TPV converter is investigated. The calculation employed fixed design parameters of $L$ = 5 mm, $T_{\text{H}}$ = 1000 K. In Fig.\ \ref{fig:5}(a), CE with varying $t_{\text{si}}$ within the range of 10 to 1000 $\mu$m is shown at three different vacuum gaps of $d$ = 10, 100, and 1000 nm. $d$ = 1000 nm represents the far-to-near field transition regime while $d$ = 10 nm and 100 nm represent the near-field regime \cite{song2022thermophotovoltaic}. CE of the TPV converter with Au reflector is shown together for comparison. In this case, the Si intermediate layer does not serve as a thermal conductor layer since the PV cell can be kept at $T_{\text{cool}}$ as cooled from the bottom of the backside reflector. Therefore, CE monotonically decreases with increasing $t_{\text{si}}$ regardless of $d$. It is because 10 $\mu$m is already enough thickness for Si intermediate layer to prevent the parasitic absorption loss caused by surface mode heat transfer (see Fig.\ \ref{fig:3}(a)) and further increased $t_{\text{si}}$ only causes parasitic absorption by Si intermediate layer. On the other hand, Si intermediate layers serve as thermal conductor layers in bifacial TPV converters to transfer the excess heat toward cooling channels. Depending on the amount of excess heat, the optimal value for $t_{\text{si}}$ exists such that CE is maximized. When $t_{\text{si}}$ is increased beyond this value, parasitic absorption by Si intermediate layer cancels out the benefit of improved cooling of PV cell, and CE decreases in response.

When $d$ = 1000 nm, near-field radiation plays a minor role in total radiative heat transfer. The excess heat by photonic and electrical losses are relatively smaller, which means the cooling load and required conduction cross-sectional area ($\sim$2$\times$$t_{\text{si}}$) are low. Accordingly, optimal $t_{\text{si}}$ is around 20 $\mu$m for the case. When $d$ = 100 nm, both POD and excess heat largely increase due to the engagement of near-field radiation. The optimal $t_{\text{si}}$ increases by 5 times to 100 $\mu$m. If $t_{\text{si}}$ is 10 $\mu$m, temperature rise in the PV cell cancels out the benefit of bifacial NF-TPV and CE gets even lower than that of TPV converter with Au reflector at $d$ = 1000 nm. At optimal $t_{\text{si}}$ = 100 $\mu$m, although absorption by Si intermediate layer is not negligible as observed in Fig.\ \ref{fig:3}(b), CE is still larger by 2.7\% point compared to the TPV converter with Au reflector with $d$ = 100 nm and $t_{\text{si}}$ = 10 $\mu$m. When $d$ = 10 nm in the bifacial NF-TPV converter, the excess heat becomes very large so that the TPV converter cannot produce any electrical power when $t_{\text{si}}$ is smaller than 15 $\mu$m. The optimal $t_{\text{si}}$ is around 300 $\mu$m where the corresponding maximum CE is as high as 30.0\%. However, when compared to the TPV converter with Au reflector with $d$ = 10 nm and $t_{\text{si}}$ = 10 $\mu$m, the CE increment is only by 1.2\% point which is a much smaller increment than in the $d$ = 100 nm case. Therefore, $d$ = 100 nm is not only more feasible but also a better vacuum gap to emphasize the performance enhancement of bifacial NF-TPV converter over the conventional NF-TPV converters.

Fig.\ \ref{fig:5}(b) shows the CE of the bifacial TPV converter with respect to $t_{\text{si}}$ and $d$. The grey area where both $t_{\text{si}}$ and $d$ are small depicts the region where the center temperature $T_{\text{L}}$($x$=$L$/2) exceeds 600 K, where we assumed the calculation is not needed due to low CE. As in Fig.\ \ref{fig:5}(a), optimal $t_{\text{si}}$ shifting toward a larger value with decreasing $d$ can be observed. In the vacuum gap near 1000 nm, there exists oscillatory behavior of CE regardless of $t_{\text{si}}$. In the far-to-near-field transition regime, destructive interference of propagating waves near the bandgap frequency can deteriorate the absorbed above-bandgap radiation \cite{zhang2007nanomicro, basu2016near, song2022thermophotovoltaic}. Such coherence in radiative heat flux absorption leads to performance reduction at the vacuum gap near 1000 nm.

Fig.\ \ref{fig:5}(c) shows the POD of the bifacial TPV converter with respect to $t_{\text{si}}$ and $d$. POD is little affected by the $t_{\text{si}}$ when $t_{\text{si}}$ is larger than the optimal value (i.e., when PV cell cooling is sufficient for all $d$), proving that the absorbed radiation by the InAs cell is barely disturbed as $t_{\text{si}}$ is increased. Therefore, CE reduction at larger $t_{\text{si}}$ in Fig.\ \ref{fig:5}(a) primarily originates from the radiation absorption by Si intermediate layer. As $t_{\text{si}}$ is decreased below optimal values in the near-field regime ($d$$\leq$1000 nm), PV cell temperature rise causes large electrical loss which lowers the POD significantly. Coherence in absorbed above-bandgap radiation can also be observed by the decreased POD at around $d$ = 1000 nm. By the assist of constructive interference of propagating wave, the same level of POD = 0.4 W/m$^2$ can be obtained at $d$ = 650 nm and $d$ = 1050 nm, when $t_{\text{si}}$ = 100 $\mu$m. Concerning that effort to achieve a smaller vacuum gap is substantial, such coherence effect could be utilized in bifacial NF-TPV converters to achieve the same level of POD at relatively larger $d$.

In Fig.\ \ref{fig:5}(c), two additional factors for design parameter selection are overlaid on the contour. View factor $F_{\text{e}}$ expressed with solid line contour denotes the 2-dimensional view factor of one emitter for another emitter. When both emitters have lateral size $L$ and are separated approximately by 2$\times$$t_{\text{si}}$, the expression for $F_{\text{e}}$ is \cite{incropera1996fundamentals}
\begin{equation}
F_{\text{e}}=\{(1+(L/2t_{\text{si}})^2\}^{\text{1/2}}-L/2t_{\text{si}}
\label{eq:4}
\end{equation}
The value represents the proportion of diffuse radiation escaping one surface that reaches another surface. The view factor ($F_{\text{e}}$) between two emitters is calculated to quantify the amount of sub-bandgap radiative heat flux that successfully reaches the emitter at the opposite side and is recycled \cite{inoue2021integrated}. Although it needs a more complex calculation for a derivation of the rigorous proportion of the recycled radiation, we use calculated $F_{\text{e}}$ $\geq$ 0.9 as an acceptable value for simplicity. When $L$ = 5 mm, as used in Fig.\ \ref{fig:5}, $t_{\text{si}}$ $\leq$ 250 $\mu$m needs to be used regarding the $F_{\text{e}}$ limit.

Another important factor is the heat flux required at the edges of the absorbing composite for cooling which is expressed in a dashed line contour in Fig.\ \ref{fig:5}(c). It is calculated by the sum of excess heat in the absorbing composite within $x$ = 0 $\sim$ $L$/2 by the cross-sectional area of the absorbing composite (refer to Fig.\ \ref{fig:2} (b) for coordinates). The calculated $q_{\text{edge}}$ is larger at smaller $d$ where thermal loss is amplified while it decreases as $t_{\text{si}}$ is increased. Due to the need for cooling the absorbing composite at its narrow side edges, the required level of heat flux may prove unattainable. Two-phase cooling techniques that incorporate jet impingement with micro-channels \cite{sung2009chf} or liquid metal \cite{deng2022two} can safely achieve the escaped heat flux of around 200 W/cm$^2$. Therefore, we only accept the combination of design parameters that demands $q_{\text{edge}}$ smaller than 200 W/cm$^2$.

In Fig.\ \ref{fig:5}(c), the regime of $t_{\text{si}}$ and $d$ that satisfies both $q_{\text{edge}}$$\leq$200 W/cm$^2$ and $F_{\text{e}}$ $\geq$0.9 is denoted with a green dotted line boundary. Smaller vacuum gaps provide very high POD in bifacial NF-TPV converters. However, the appropriate range for Si intermediate layer thickness becomes narrower because increased cooling load demands larger $t_{\text{si}}$. Furthermore, although we defined 200 W/cm$^2$ as attainable cooling heat flux, cutting-edge two-phase cooling may be challenging to apply to our converter without deterioration. Therefore, a vacuum gap as small as 10 nm which always requires $q_{\text{edge}}$ larger than 100 W/cm$^2$ may be avoided in practical design. To demonstrate a practical converter design, a point where ($d$, $t_{\text{si}}$) = (100 nm, 100 $\mu$m) is denoted in yellow stars in all three panels of Fig.\ \ref{fig:5}. A bifacial NF-TPV converter with a combination of design parameters at the point gives CE of 26.8\% and POD of 4.38 W/cm$^2$ with 1000 K emitters and single-junction InAs cell. The design gives $F_{\text{e}}$ = 0.91 and $q_{\text{edge}}$ = 63 W/cm$^2$ which are both in their acceptable ranges. Here, the average temperature of the absorbing composite is 308.3K which is only 8.3 K higher than the cooling temperature.

\subsection{Effect of emitter temperature and Si intermediate layer thickness}

\begin{figure}[!t]
    \centering
    \includegraphics[width=0.7\textwidth]{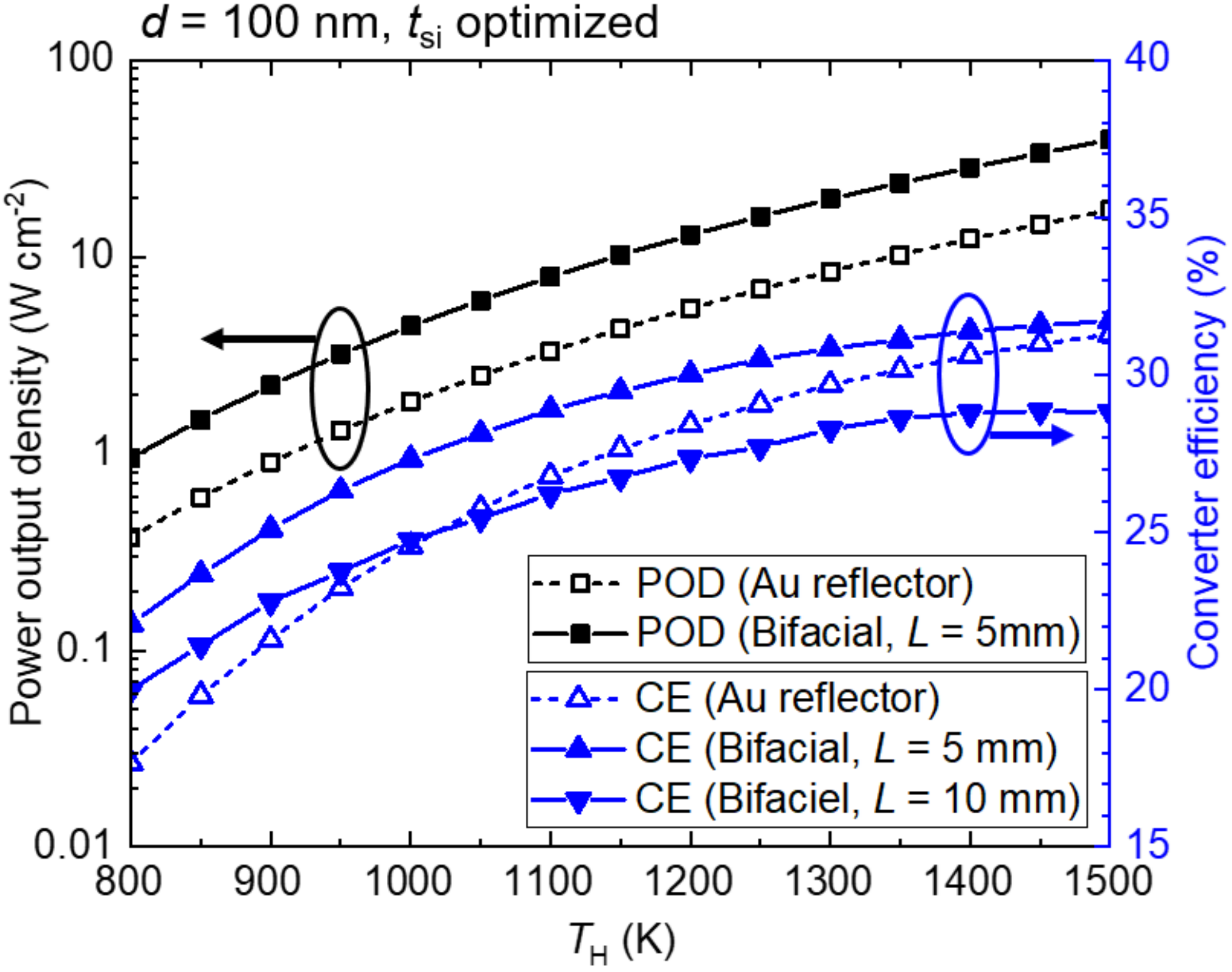}
    \caption{Power output density and converter efficiency of bifacial NF-TPV converter and NF-TPV converter with Au reflector with respect to emitter temperature, where the thickness of the Si intermediate layer is optimized for each respective emitter temperature.}
    \label{fig:6}
\end{figure}

In TPV converters, it is frequently beneficial to produce the same POD from lower emitter temperatures. Lower emitter temperature increases both the cost-effectiveness of the TPV converter and the availability of the thermal source. Bifacial NF-TPV converters have the potential to enhance POD compared to conventional NF-TPV converters with Au reflectors (Au reflector case) at a wide range of emitter temperatures. In Fig.\ \ref{fig:6}, POD and CE are compared for two NF-TPV converters at emitter temperatures ranging from 800 to 1500 K. In calculation, $d$ = 100 nm was used for consistency. Different $t_{\text{si}}$ was selected for each $T_{\text{H}}$ within the range of 10 $\sim$ 1000 $\mu$m so that CE can be maximized, based on the trade-off described in Fig.\ \ref{fig:5}(a). For the Au reflector case, $t_{\text{si}}$=10 $\mu$m was used at all $T_{\text{H}}$.

In bifacial NF-TPV converters, increment in POD exists for every $T_{\text{H}}$, varying within the enhancement of 2.28 times at $T_{\text{H}}$ = 1500 K to 2.52 times at $T_{\text{H}}$ = 800 K compared to the Au reflector case. Such enhancement enables a much smaller PV cell area for a certain electrical power to reduce the material and operating costs. Alternatively, it enables lower $T_{\text{H}}$ required for a certain POD. To be specific, the required $T_{\text{H}}$ to achieve POD of 1 W/cm$^2$ decreases by 108 K from $T_{\text{H,Au}}$ = 913 K to $T_{\text{H,Bi}}$ = 805 K where subscripts `Au' and `Bi' mean NF-TPV converter with Au reflector and bifacial NF-TPV converter, respectively. For achieving 9 W/cm$^2$, the required $T_{\text{H}}$ decreases by 192 K from $T_{\text{H,Au}}$ = 1317 K to $T_{\text{H,Bi}}$ = 1125 K. Consequently, the required electrical power can be obtained from lower emitter temperatures for the bifacial NF-TPV converters, thereby allowing for a broader range of thermal sources to be utilized. When employing phase-change material-based thermal energy storage as the heat source for the TPV converter, an emitter temperature of $T_{\text{H}}$ = 1317 K can be exclusively attained utilizing MgF$_{\text{2}}$ as the phase-change material. In contrast, $T_{\text{H}}$ = 1125 K can be attained using various phase-change materials, including Na$_{\text{2}}$CO$_{\text{3}}$, K$_{\text{2}}$Co$_{\text{3}}$, KF, and NaF \cite{pielichowska2014phase}.

Comparing the converter efficiencies of the bifacial NF-TPV converter and Au reflector case in Fig.\ \ref{fig:6}, there exists an increment at the whole range of $T_{\text{H}}$ calculated. At $L$ = 5 mm, the increment is 4.4\% point at $T_{\text{H}}$=800 K and 0.4\% point at $T_{\text{H}}$ = 1500 K. The increment in CE is prominent at lower $T_{\text{H}}$, which makes bifacial NF-TPV an attractive choice for electrical energy generation from mid-temperature range thermal sources. However, when $L$ is doubled to 10 mm, the maximum achievable CE of the bifacial NF-TPV converter is decreased by 2.1\% when $T_{\text{H}}$ = 800 K and by 2.9\% when $T_{\text{H}}$ = 1500 K. As a result, $T_{\text{H}}$ larger than 1017 K always causes lower CE in bifacial NF-TPV converters compared to the Au reflector case. It is mainly because a larger lateral size increases the average and maximum temperature of the absorbing composite which, in turn, causes larger electrical loss by non-radiative recombination.

\begin{figure}[!t]
    \centering
    \includegraphics[width=1\textwidth]{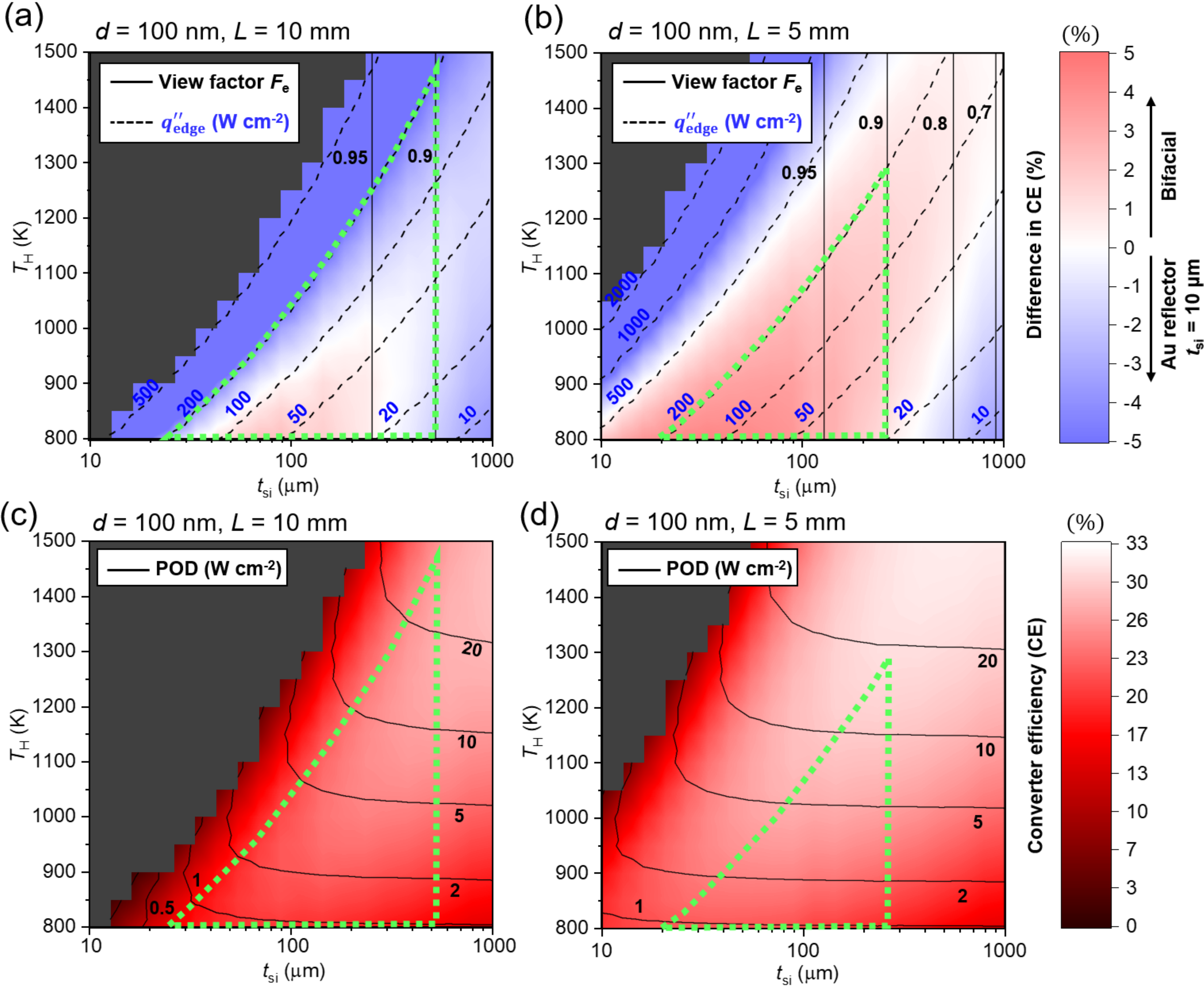}
    \caption{(a), (b) Difference in converter efficiency (CE) of the bifacial NF-TPV converter compared to the NF-TPV converter with Au reflector with respect to $t_{\text{si}}$ and $T_{\text{H}}$, together with the view factor between two emitters and the edge cooling heat flux, (c), (d) CE of the bifacial NF-TPV converter with respect to $t_{\text{si}}$ and $T_{\text{H}}$, together with power output density information.}
    \label{fig:7}
\end{figure}

To additionally discuss the effect of $t_{\text{si}}$ and $T_{\text{H}}$, the CE differences between the two NF-TPV converters are plotted in Fig.\ \ref{fig:7}(a) and Fig.\ \ref{fig:7}(b) calculated with $L$ = 10 mm and $L$ = 5 mm, respectively. The red color shows the region where the CE of the bifacial NF-TPV converter is greater than that of the Au reflector case with $t_{\text{si}}$ = 10 $\mu$m and the blue color shows the opposite. The grey area depicts the region where the center temperature $T_{\text{L}}$($x$=$L$/2) exceeds 600 K. Similar to Fig.\ \ref{fig:5}(c), the region for $q_{\text{edge}}$ $\leq$ 200 W/cm$^2$ and $F_{\text{e}}$ $\geq$ 0.9 is depicted with green dotted line boundary. In Fig.\ \ref{fig:7}(a) where $L$ = 10 mm, the red region exists only in the lower $T_{\text{H}}$ below 1017 K and at $t_{\text{si}}$ around 100 um. Due to the PV cell temperature rise and more electrical loss, CE exceeding that of NF-TPV with Au reflector is challenging at higher emitter temperatures. In Fig.\ \ref{fig:7}(b) where $L$ = 5 mm, the red region stretches up to $T_{\text{H}}$ = 1500 K indicating larger CE than Au reflector case is possible at higher emitter temperatures. However, no region with $T_{\text{H}}$ $\geq$ 1290 K can meet the requirements for the view factor because of the shorter $L$ (refer to Eq.\ (3)). In Fig.\ \ref{fig:7}(c) and Fig.\ \ref{fig:7}(d), CE of bifacial NF-TPV converter is plotted with $L$ = 10 mm and $L$ = 5 mm, respectively. POD information is overlaid together with the regions that satisfy requirements for $q_{\text{edge}}$ and $F_{\text{e}}$ already shown in Fig.\ \ref{fig:7}(a) and Fig.\ \ref{fig:7}(b). It is seen that $L$ = 5 mm case shows larger CE at the same combination of ($T_{\text{H}}$, $t_{\text{si}}$) compared to $L$ = 10 mm case. Accordingly, the calculable region (i.e., where maximum PV cell temperature is lower than 600 K) is broader in the $L$ = 5 mm case. However, since the lateral size is smaller, $t_{\text{si}}$ that corresponds to $F_{\text{e}}$ = 0.9 is smaller. Therefore, thinner Si intermediate layers should be used and the allowable emitter temperature becomes limited to 1290 K in which the maximum POD is given as 19.0 W/cm$^2$ at the vertex. To achieve an even larger POD, $L$ has to be larger so that thicker $t_{\text{si}}$ is allowable that can suffice the conduction cross section required. In Fig.\ \ref{fig:7}(c), the region where requirements for $q_{\text{edge}}$ and VF are met stretches up to 1470 K where $t_{\text{si}}$ = 520 $\mu$m at the vertex. Although the CE is lower by 3.5\% in comparison to the NF-TPV converter with Au reflector at the vertex, it can provide 29.3 W/cm$^2$ of POD which is larger by 2.01 times than that of Au reflector case at the same $T_{\text{H}}$.

\subsection{Effect of lateral size and Si intermediate layer thickness}

\begin{figure}[!t]
    \centering
    \includegraphics[width=1\textwidth]{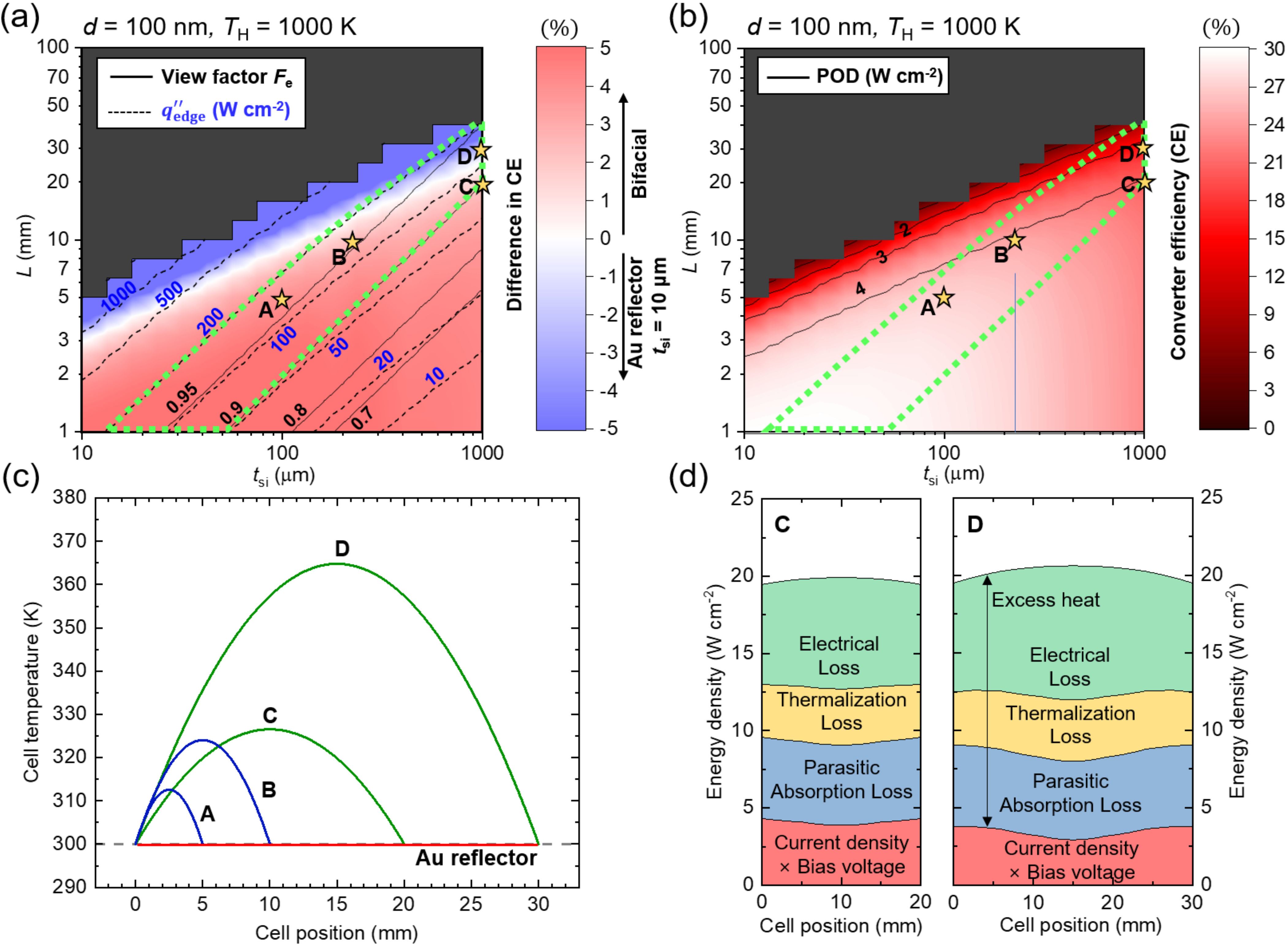}
    \caption{(a) Difference in converter efficiency (CE) of the bifacial NF-TPV converter compared to the NF-TPV converter with Au reflector with respect to $t_{\text{si}}$ and $L$ together with the view factor between two emitters and the edge cooling heat flux, (b) CE of the bifacial NF-TPV converter with respect to $t_{\text{si}}$ and $L$, together with power output density information. (c) The lateral temperature profiles of the absorbing composites with different values of design parameters denoted as points $A \sim D$ in (a) and (b), (d) Lateral position-dependent loss analysis for points $C$ and $D$.}
    \label{fig:8}
\end{figure}

In Fig.\ \ref{fig:8}, the performance of bifacial NF-TPV converters is analyzed with different $L$ and $t_{\text{si}}$ values at fixed $T_{\text{H}}$ of 1000 K. The difference in CE compared to that of the Au reflector case with $t_{\text{si}}$ = 10 $\mu$m is plotted in Fig.\ \ref{fig:8}(a). Calculated $q_{\text{edge}}$ and $F_{\text{e}}$ are overlaid on the plot. Region for $q_{\text{edge}}$ $\leq$ 200 W/cm$^2$ and $F_{\text{e}}$ $\geq$ 0.9 is depicted with a green dotted line boundary, and points with four different [$L$, $t_{\text{si}}$] combinations $A \sim D$ are denoted with yellow stars. In Fig.\ \ref{fig:8}(b), CE of bifacial NF-TPV is plotted with the same coordinate as in Fig.\ \ref{fig:8}(a) together with POD information and allowable region regarding $q_{\text{edge}}$ and $F_{\text{e}}$. As the lateral size of the bifacial NF-TPV converter increases, the acceptable range for $t_{\text{si}}$ progressively shifts towards greater values of $t_{\text{si}}$. At the same time, the maximum achievable CE and POD decreases. When $L$ = 5 mm, $t_{\text{si}}$ between 70 $\mu$m and 250 $\mu$m is allowed where CE difference with that of Au reflector case is maximum at point $A$ where $t_{\text{si}}$ = 100 $\mu$m to be 2.7\%. At point $B$ where $L$ = 10 $\mu$m and $t_{\text{si}}$ = 220 $\mu$m, converter efficiencies of two NF-TPV converters become similar and the difference is by 0.2\%. However, a POD of 4.15 W/cm$^2$ is achievable in the bifacial NF-TPV converter which is 2.29 times larger than that of the Au reflector case. When $L$ = 20 mm, maximum CE is reached at point $C$ where $t_{\text{si}}$ = 1000 $\mu$m, where 4.0\% smaller CE of 20.6\% but 2.24 times larger POD of 4.06W/cm$^2$ is achieved compared to the Au reflector case. When $L$ is further increased by 1.5 times to become 30 mm (point $D$), CE is further lowered to 16.8\% where POD is also much reduced to 3.41 W/cm$^2$ which is only 1.88 times that of the Au reflector case.

In Fig.\ \ref{fig:8}(c), the lateral temperature profile of absorbing composite $T_{\text{L}}$($x$) for 4 points $A$, $B$, $C$, and $D$ marked in Fig.\ \ref{fig:8}(a) and Fig.\ \ref{fig:8}(b) are plotted together. In point $A$, the maximum temperature rise from the cooling temperature 300 K is by 13 K. Therefore, the increase in the electrical loss by temperature rise is minor, i.e., cooling in the lateral direction is hardly a problem when $L$ = 5 mm, although parasitic absorption loss may have increased by the existence of the 100-um-thick Si intermediate layer. In point $B$ where $L$ is doubled to 10 mm, the distance from the edge to the center is doubled while the required cooling heat flux at the edge is also doubled. As a result, although $t_{\text{si}}$ has increased by 2.2 times, the maximum temperature rise from $T_{\text{cool}}$ becomes much greater than in point $A$ to be 24 K. Hence, the more heat flux absorbed in the PV cell is lost as turning into the electrical loss, and both POD and CE are decreased compared to point $A$. In point $C$, $L$ is doubled from point $B$ to be 20 mm. However, $t_{\text{si}}$ was also increased by more than 4 times to be 1000 $\mu$m. As a result, the maximum temperature rise is 26 K, a value similar to that in point $B$. Since the amounts of electrical loss in points $B$ and $C$ are similar, their power output densities are also similar. However, 4 times thicker $t_{\text{si}}$ causes enhanced absorption to the Si intermediate layer. As a result, CE is lowered by 4.2\% in point $C$ compared to point $B$. In point $D$, $L$ is increased by 1.5 times but the maximum PV cell temperature becomes 364 K which, primarily by the electrical loss, lowers the POD by 0.64 W/cm$^2$ and CE by 3.7\% compared to point $C$.

In Fig.\ \ref{fig:8}(d), electrical loss, thermalization loss and parasitic absorption loss, and current density multiplied by bias voltage at the maximum power point (i.e., local POD) are shown with respect to lateral cell position. Their spatial average values represent electrical losses, photonic losses, and power output densities, respectively. At both points $C$ and $D$, an increase in temperature at the center region results in a corresponding increase in electrical loss, leading to a reduction in current density at that location. The thermalization loss and parasitic absorption loss terms are relatively unchanged in the two cases. In other words, the temperature change in absorbing composite mainly affects electrical loss rather than photonic losses such as radiative recombination. On the other hand, the sum of all four components indicates the absorbed radiative heat flux to the absorbing composite. At the center region, the bandgap energy of the PV cell decreases with its temperature increase which leads to the increment of the absorbed radiative heat flux at that location. However, the additional radiation absorption is canceled out by the increase in electrical loss and local POD becomes lower at the center region of the PV cell.

As seen in Fig.\ \ref{fig:8}, the lateral size of the bifacial NF-TPV converter is limited to 20 mm when trying to achieve either enhancement in CE or more than two-fold enhancement in POD compared to the NF-TPV converter with the Au reflector. However, it is worth mentioning that achieving a vacuum gap of 100 nm with an emitter with large width is incredibly challenging. For the near-field radiation experiments, the maximum width of the thermal emitter and receiver that has maintained a vacuum gap smaller than 200 nm is 10 mm up to date \cite{watjen2016near, ying2019super}. Likely, a single bifacial NF-TPV converter will possess a lateral size no larger than 10 mm where the enhancement in CE and 2.29 times larger POD can be simultaneously achieved compared to a conventional NF-TPV converter with Au reflector. Furthermore, as seen in Fig.\ \ref{fig:7}, the bifacial NF-TPV converter becomes a more attractive option at lower emitter temperatures between 800 K and 1000 K. At lower emitter temperatures, regulation for $L$ and $t_{\text{si}}$ is much relieved while the relative performance increment gets greater. Consequently, the required emitter temperature for a certain POD is decreased by more than 100 K when incorporating bifacial TPV conversion to NF-TPV conversion (refer to Fig.\ \ref{fig:6}). Therefore, we conclude that the suggested bifacial NF-TPV converter can provide a more cost-effective and energy-efficient TPV conversion compared to existing configurations for TPV conversion.

\section{Conclusion}\label{conclusion}
Thermophotovoltaic conversion has great potential for eco-friendly and high-performance electrical power generators. The desire for greater CE and POD in TPV converters has led to the development of NF-TPV converters and a recently suggested bifacial TPV converter. NF-TPV converters utilize radiative heat flux increment by photon tunneling while bifacial TPV converters provide near-perfect sub-bandgap radiation recycling. In this study, we successfully accumulated the benefits of the NF-TPV converter and bifacial TPV converter despite the risk of PV cell temperature rise coming from the requirement to release large excess heat at the side edges. Key design fundamentals included the employment of transparent Si intermediate layers that carry heat by thermal conduction and suppress the absorption of sub-bandgap surface mode radiation in PV cells. We developed an opto-electro-thermal calculation algorithm to iteratively find the numerical solution for the performance of the TPV converter. A combination of design parameters that enable simultaneous enhancement in POD and CE compared to conventional NF-TPV converter were explored. We could provide practical design criteria for the bifacial NF-TPV converter which involves consideration of the required heat flux at the side edges and the view factor between emitters. The proposed designs have been verified to have feasible vacuum gap sizes and emitter areas comparable to those achieved in state-of-the-art near-field thermal radiation experiments. Therefore, we assume bifacial NF-TPV converter with significant performance enhancement can be experimentally achieved with the current technology level. Owing to a greater than two-fold enhancement in the POD compared to NF-TPV converters, the necessary emitter temperature for achieving a specified POD can be substantially decreased, facilitating more economical power generation. Furthermore, we believe that the performance of the bifacial NF-TPV converter can be further boosted when employing multi-junction PV cells. Our findings will pave the way for viable and high-performance TPV conversion, thus contributing to the advancement of this promising energy conversion technology.

\section{Acknowledgements}\label{acknowledgements}
This work is supported by the Basic Science Research Program through the National Research Foundation of Korea Grants funded by the Ministry of Science and ICT under Grant NRF-2019R1A2C2003605 and Grant NRF-2022R1C1C2008309.

\section{Declaration of interests}\label{Declaration of interests}
The authors declare that they have no known competing financial interests or personal relationships that could have appeared to influence the work reported in this paper.



\bibliographystyle{elsarticle-num-names}
\bibliography{mainBib}







\end{document}